# A semiconductor source of triggered entangled photon pairs?

A. Gilchrist, K. J. Resch, and A. G. White

*Department of Physics and Centre for Quantum Computer Technology,*

*University of Queensland, Brisbane QLD 4072, Australia*

**The realisation of a triggered entangled photon source will be of great importance in quantum information, including for quantum key distribution and quantum computation. We show here that: 1) the source reported in "A semiconductor source of triggered entangled photon pairs"[1] is not entangled; 2) the entanglement indicators used in Ref.[1] are inappropriate, relying on assumptions invalidated by their own data; and 3) even after simulating subtraction of the significant quantity of background noise, their source has insignificant entanglement.**

Stevenson et al.[1] produce pairs of photons via radiative decay of biexcitons in single quantum dots, and measure strong coherences between two polarisation populations of the photons, a necessary but not sufficient condition for entanglement. Establishing the degree of entanglement between two qubits is a solved problem[2–6]: Ref.[1] apply none of these standard techniques.

Without loss of generality[6], we use the *tangle*, $0 \leq T \leq 1$, to unambiguously quantify the degree of entanglement: $T=0$ means there is no entanglement (any correlations are classical); $T=1$ means that the state is maximally entangled[4]. The tangle can be directly calculated from the two-photon density matrix obtained via quantum state tomography[8]. For all the density matrices measured by Stevenson et al.[1] there is strictly no entanglement, as $T=0$[9]. Note that this is true even given the large coherences measured for the quantum dots B, C in Figs 3b), d).

In Fig. 2c) Stevenson et al.[1] measure the *degree of correlation*[11], for quantum dots A & B. Entanglement is indicated by a constant value *greater* than $|50|\%$[12]. The authors arbitrarily fit a horizontal line to the data for dot B[13], yielding a value of $(22.2\pm2.8)\%$. Regardless of the choice of fitted curve, the fact that the mean is $9.9\sigma$ *less* than 50% means this data provides no evidence of entanglement: consistent with $T=0$ as measured above. This highlights that the shape of the degree of correlation curve is a very poor entanglement indicator, since it varies greatly even between different unentangled sources, e.g. quantum dots A & B in Fig. 2c)[1].

Stevenson et al.[1] also gauge entanglement by applying an eigenvalue method. Entanglement is indicated by the largest eigenvalue of the two-photon density matrix exceeding 0.5, *only* if the individual photons are unpolarised. Self-consistency demands that the same data used to gauge entanglement is also used to check the photon polarisation. We obtain the polarisation for each single photon from Stevenson et al.'s two-photon density matrix by tracing out the other photon, and find one photon is partially-polarised, with a *degree of polarisation*[14–16] of (4.5±1.9)%, rendering the eigenvalue method invalid in this case. Note that the other photon has zero polarisation, (0±1.1)%, due to the artificial normalisation imposed in Ref.[1] (see Methods).

Based on the data in both Fig. 2 and Fig. 3 Stevenson et al.[1] conclude that "the measurements presented above clearly suggest that dots with small exciton splitting emit entangled photons". As we have shown, this conclusion is not supported by their data or methods.

An entangled photon source will be degraded by unwanted background light. Stevenson et al.[1] identify 49% (no error given) of photon pairs as arising from "background due to dark counts and emission from layers other than the dot". They simulated an improved source by subtracting the projected (as opposed to directly measured) number of background counts, but did not show the resulting density matrices or data. To ascertain the entanglement of this simulated source we subtracted a component representing 49% unpolarised light from the density matrix of Fig. 3d) (see Methods). Even after the simulated removal of this significant amount of noise, the density matrix has insignificant tangle, $T=0.028 \pm 0.022$[17]. Naturally, removing unpolarised background from a partially-polarised source serves to further increase the degree of polarisation of the partially-polarised photon, to (8.8±3.4)%: the eigenvalue method remains invalid after background subtraction.

In conclusion, Stevenson et al.[1] have not satisfied the standard of proof required to claim "A semiconductor source of triggered entangled photon pairs".

**Methods**

The probability data in Table 1[18] was used to tomographically generate Figure 3d) in Ref.[1]. Standard tomography uses complete measurement sets to ensure that the sum of the probabilities is unity, e.g. $P_{VV}+P_{VH}+P_{HV}+P_{HH}=1$. Repeating this in each measurement basis allows for compensation of measurement drift. This was not done by Stevenson et al., who instead used an artificial normalisation that set the sum of each of their first four probability pairs to a half, e.g. $P_{VV}+P_{VH}=0.5$. The effect of Stevenson et al.'s extra constraint is to *force* one photon in each pair to be unpolarised.

The probabilities, $P$, were calculated from the the second-order correlations, $g^{(2)}(\tau=0)$, between photons 1 & 2: $P=\frac{1}{2}g_A^{(2)}/(g_A^{(2)}+g_B^{(2)})$, where A & B refer to orthogonal measurement settings on photon 2[18]. It is this formula that imposes the additional normalisation constraint described above. In turn, the second-order correlation was given by $g^{(2)}=C/(N/n)$, where $C$ is the number of pairs detected in the same laser cycle, $N/n$ is the average number of pairs in different laser cycles, and $n$ is the number of measured finite delay peaks. The error is given by $\Delta g^{(2)}=g^{(2)}\sqrt{1/C+1/N}$, which assumes that $\Delta n=0$ and that the errors in $C$ and $N$ are Poissonian, i.e. $\Delta C=\sqrt{C}$, $\Delta N=\sqrt{N}$.

To perform unbiased quantum state tomography and a full error analysis, we requested the underlying data used to obtain the normalised values, i.e., $C$, $N$, and $n$. This data has not been provided and indeed Stevenson et al. claim this data does not exist[18]. This is curious, given many references to count data in Ref.[1] e.g. "The total number of coincident pairs detected over the course of an experiment is typically up to 1,000, which dictates the measurement errors." In further communication, Stevenson et al. say that "We stated that the number of coincidences detected over the course of an experiment was typically 1000 simply as a guideline so that others would be able to recreate our experiment successfully."[18].

This raises the issue of the uncertainties. Stevenson et al. say "Table 1 below lists the errors associated with the count rates. They are indeed based on Poissonian noise."[18] Poissonian uncertainties are defined as $\sqrt{N}$, where $N$ is the number of counts: they cannot be calculated from a probability. It is not clear how Stevenson et al. obtained the uncertainties in Table 1.

In the absence of count data, we used Stevenson et al.'s normalised values to tomographically reconstruct the density matrix,

$$\boldsymbol{\rho}_{3d} = \begin{bmatrix} 0.3261 & -0.0096 & -0.0101 & 0.1038 \\ -0.0096 & 0.1739 & 0.0009 & 0.0101 \\ -0.0101 & 0.0009 & 0.1953 & 0.0105 \\ 0.1038 & 0.0101 & 0.0105 & 0.3047 \end{bmatrix} + i \begin{bmatrix} 0 & -0.0135 & 0.0166 & 0.0002 \\ 0.0135 & 0 & -0.0235 & -0.0166 \\ -0.0166 & 0.0235 & 0 & 0.0202 \\ -0.0002 & 0.0166 & -0.0202 & 0 \end{bmatrix},$$

which agrees with that provided by Stevenson et al[18]. For all other density matrices in Figure 3 of Ref.[1], we estimated the values of the elements directly from the diagrams, assuming that the imaginary parts were zero. The linear entropy and tangle are calculated directly from the reconstructed and estimated density matrices, as described in Refs[3–6]. In order to estimate uncertainties in these derived quantities, an ensemble of 5000 density matrices was generated by: 1) creating a new data set by sampling from a Gaussian distribution centred on the mean values supplied by Stevenson et al., with standard deviation equal to the supplied errors and 2) applying maximum-likelihood tomography to each such data set. The tangle, and the degree of polarisation of each photon, were calculated for each of these 5000 density matrices. Our uncertainties are the standard deviation of these quantities.

Subtraction of 49% unpolarised background light (no error given) was modelled by subtracting 0.49 $\boldsymbol{I}/4$ from a large ensemble of density matrices constructed as above, and renormalising appropriately, where $\boldsymbol{I}$ is the 4×4 identity matrix. This was done until 5000 physical density matrices were obtained (all non-negative eigenvalues), and then uncertainties were calculated as above.

In summary, Stevenson et al. have: supplied insufficient experimental data to allow reproduction or verification of their error calculations; used an artificial normalisation method that biases their reconstructed states; and supplied no error for their estimate of the background noise.

Table 1: Probabilities for Fig. 3d[1]

| Measure | $P$ | $\Delta P$ | Measure | $P$ | $\Delta P$ | Measure | $P$ | $\Delta P$ | Measure | $P$ | $\Delta P$ |
|---|---|---|---|---|---|---|---|---|---|---|---|
| VV | 0.30470 | 0.00606 | LV | 0.25592 | 0.00737 | DL | 0.23479 | 0.00919 | HD | 0.24041 | 0.01065 |
| VH | 0.19530 | 0.00496 | LH | 0.24408 | 0.00562 | DD | 0.30285 | 0.00463 | HR | 0.23647 | 0.01235 |
| HH | 0.32605 | 0.01084 | DH | 0.25058 | 0.00459 | LD | 0.26214 | 0.00631 | VR | 0.27020 | 0.00773 |
| HV | 0.17395 | 0.01409 | DV | 0.24942 | 0.00651 | VD | 0.26054 | 0.00511 | LR | 0.30477 | 0.00769 |

---

[1] Stevenson, R. M., Young, R. J., Atkinson, P., Cooper, K., Ritchie, D. A. & Shields, A. J. *Nature* **439**, 179-182 (2006).

[2] Bennett C. H., DiVincenzo, D. P., Smolin, J. A., & Wootters W. K. *Physical Review A* **54**, 3824 (1996).

[3] Wootters, W. K. *Physical Review Letters* **80**, 2245 (1998).

[4] Coffman, V., Kundu, J. & Wootters, W. K. *Physical Review A* **61**, 052306 (2000).

[5] White, A. G., James, D. F. V., Munro, W. J. & Kwiat, P. G. *Physical Review A* **65**, 012314 (2001).

[6] The entanglement of formation[2], $E_F$, concurrence[3,4], $C$, and tangle[5], $T$, are directly related, and are *quantitative*, measuring the degree of entanglement. The tangle is the most conservative of the three metrics, and satisfies tight inequalities not shared by the others.[4] Other methods, such as the Peres condition[7], or fidelity with a maximally-entangled state, are *qualitative*, indicating but not measuring entanglement. Note: $C=\sqrt{T}$; $E_F = h((1+\sqrt{1-T})/2)$, where $h(x) = -x \log_2 x - (1-x)\log_2(1-x)$.

[7] Peres, A. *Physical Review Letters* **77**, 1413 (1996).

[8] James, D. F. V., Kwiat, P. G., Munro, W. J. & White, A. G. *Physical Review A* **64**, 052312 (2001).

[9] The amount of entanglement is limited by the purity of the state, which can be quantified by the *linear entropy*, $S_L$, calculated from the density matrix: $S_L=0$ for pure, highly-ordered states; $S_L=1$ for maximally mixed states[5]. States with $S_L > 8/9$ always have zero entanglement[10]: the states in Fig. 3[1] are highly mixed, $0.92 \leq S_L \leq 0.99$, and so the tangle is necessarily zero, consistent with the direct calculations.

[10] Munro, W. J., James, D. F. V., White, A. G. & Kwiat, P. G. *Physical Review A* **64** 030302 (2001).

[11] The degree of correlation is $\mathcal{C}(\theta,\theta) = (N(\theta,\theta) - N(\theta,\theta+\pi/2))/(N(\theta,\theta) + N(\theta,\theta+\pi/2))$, where $-1 \leq \mathcal{C} \leq 1$ and $N(\theta_1,\theta_2)$ is the coincidence count rate with linear polarisers analysing photon 1 (2) at $\theta_1$ ($\theta_2$). For the maximally entangled state $(|HH\rangle + |VV\rangle)/\sqrt{2}$, $\mathcal{C}(\theta,\theta)=1$; for mixed (unpolarised) states, $\mathcal{C}(\theta,\theta)=0$.

[12] Young, R. J., Stevenson, R. M., Atkinson, P., Cooper, K., Ritchie, D. A., & Shields, A. J. *New Journal of Physics* **8**, 29 (2006).

[13] Stevenson et al.[1] chose to fit a sinusoid for dot A, and a constant value for dot B. We note that for dot B, with better statistical significance, one can fit a sinusoidal function with the same period as dot A that

ranges from 20% to 25% (reduced chi-squared of $\chi^2$=0.065 c.f $\chi^2$=0.13). There are no statistical grounds for asserting that "...the degree of correlation for degenerate dot B is independent of the measurement basis"[1].

[14] The degree of polarisation is the length of the Stokes vector[15], $P=\sqrt{1-4\,|\boldsymbol{\rho}_i|}$, where $|\boldsymbol{\rho}_i|$ is the determinant of the single polarisation density matrix, $\boldsymbol{\rho}_i$.[16]

[15] Stokes, G. G. *Transactions of the Cambridge Philosophical Society* **9**, 399 (1852).

[16] Mandel, L., and Wolf, E. *Optical Coherence and Quantum Optics.* (Cambridge University Press, Cambridge, 1995). pp. 353-354.

[17] We note that background subtraction indicates the potential of an entangled source, but that an entangled source which requires background subtraction is not useful in quantum information. For example, the security of the Ekert protocol in quantum cryptography requires actual, as opposed to virtual, entanglement.

[18] Stevenson, R. M., Young, R. J., Atkinson, P., Cooper, K., Ritchie, D. A. & Shields, A. J., *private communication* (2006).